\magnification1200


\vskip 2cm
\centerline
{\bf  E11 and the non-linear dual graviton}
\vskip 1cm
\centerline{ Alexander G. Tumanov }
\medskip
\centerline {School of Physics and Astronomy,}
\centerline{ Tel Aviv University, Ramat Aviv 69978, Israel}
\medskip\centerline {and }
\medskip
\centerline {Peter West}
\centerline{Department of Mathematics ,}
\centerline{King's College, London WC2R 2LS, UK}
\vskip 2cm
\leftline{\sl Abstract}
The non-linear  duality relation between the gravity and dual gravity fields are found in E theory by carrying out $E_{11}$ variations of previously found duality relations. We also find the  dual graviton equation of motion up to the addition of some  very specific terms whose coefficients are not determined. Using the calculations in this paper this ambiguity was resolved in reference [15] where the full non-linear dual gravity equation was found. As a result the  equations of motion in E theory have now been found at the full non-linear level up to, and including, level three, which contains the dual graviton field. When truncated to contain fields at levels three and less, and the spacetime is restricted  to be the familiar eleven dimensional space time, the equations are equivalent to those of eleven dimensional supergravity.

\vskip2cm
\noindent

\vskip .5cm

\vfill
\eject
{\bf 1 Introduction}
\medskip
It was conjectured that the non-linear realisation of the semi-direct product of $E_{11}$ with its vector representation ($l_1$), denoted $E_{11}\otimes_s l_1$, leads to a theory, called E theory,  that contains   the eleven dimensional  supergravity theory [1,2]. E theory contains an infinite number of fields that live on a space time that has an infinite number of coordinates. However,  the fields  at low levels are just those of eleven dimensional supergravity, in particular at levels zero, one and two we find the graviton, three and six form fields respectively. Furthermore, 
the lowest level coordinate is that of our familiar eleven dimensional space time. Following early efforts, for example reference [3],  it has been shown that E theory  leads to essentially unique equations of motion  which are second order in derivatives for the graviton and form fields [4,5]. When these equations of motion are restricted to contain  only the supergravity fields which live on the usual spacetime they   are just those of eleven dimensional supergravity [3,4]. By taking different decompositions of $E_{11}$ one can find all the maximal supergravity theories including the gauge supergravities and it is 
inevitable that the analogous  result for the equations of motion  holds in all dimensions, for a review and references therein see [6]. 
\par
The equations of motion at the linearised level have been found in eleven dimensions up to and including  level four in the fields [7]. The field at level three is the dual graviton while  at level four there are  three  fields.   It was shown that the dual graviton field obeyed an equation that did correctly describe the degrees of freedom of gravity and  while one of the fields at level four leads to Romans theory,  when dimensional reduced to ten dimensions,  one of the other  level four fields was dual of the three form. The degrees of freedom resulting from the full system of equations were those of eleven dimensional supergravity. 
\par
The dual graviton first appeared in five dimensions in the reference [8]. This paper was one of the first  to consider the dynamics of fields which carried  mixed symmetry indices and it constructed  the equation of motion  of the field $\phi_{ab,c }$ with the symmetries $\phi_{ab,c }= -\phi_{ba, c}$ in a general dimension. The author noted that in the massless case,  and in five dimensions, the equation of motion for this  field had the same number of on-shell degrees of freedom as the graviton. In reference [9] the field $\phi _{a_1\ldots a_{D-3}, b} $ was considered in $D$ dimensions and suggested as a candidate for the dual graviton in the sense that a quantity which contained two space-time derivatives acting on this field could be regarded as being  dual to the Riemann tensor. By assuming the existence of an  appropriate  light cone formulation, which would suitable restrict the indices on the irreducible component of this  field to take only $D-2$ values,  it was realised that it  had the correct number of degrees of freedom to describe gravity as the $[a_1\ldots a_{D-3}]$ would be  equivalent to a single index. 
\par
As we have mentioned the field that occurs in the $E_{11}\otimes_s l_1$ non-linear realisation at level three  in eleven dimensions has the indices $h _{a_1\ldots a_{8}, b} $ and it was proposed that this would satisfy a duality relation with the usual gravity field that was first order in derivatives, generalising the duality between the three and six form fields found at levels one and two respectively [1]. Indeed an explicit equation of this type,   generalised to $D$  dimensions, and  using the field $h _{a_1\ldots a_{D-3}, b} $ was given. It was shown that one could take a derivative of this duality relation  so as to obtain an equation for the gravity field or the dual gravity field [1]. As a result it was shown that the dual gravity field $h _{a_1\ldots a_{8}, b} $ did indeed correctly propagate the degrees of freedom of gravity.  In reference [10] it was shown that  the formulation of  dual gravity that resulted in five dimensions in [1] was  the same as that given earlier in reference     [8]. 
\par
Although the gravity-dual gravity relations given in  reference [1] did correctly propagate the gravity degrees of freedom this was at  the linearised level and  this relation was postulated  rather than  derived from the $E_{11}\otimes_s l_1$ non-linear realisation in that paper. Subsequently,   a no go theorem was shown in reference [11] in which it was argued that one could not construct a theory that was equivalent to Einstein's theory using the non-linear graviton field. 
\par
\par
In this paper, using the $E_{11}\otimes_s l_1$ non-linear realisation, we will find a fully non-linear,  the second order in derivatives,  equation of motion satisfied by the dual graviton up to a set of very specific terms whose coefficients are not determined. We will also find the equation, which is first order in derivatives and  that relates the gravity field to the dual gravity field. By projecting the later relation we will recover the 
equation of motion of eleven dimensional supergravity for the gravity field,  previously  derived from the non-linear realisation. We will derive these equations by carrying out the $E_{11}$ variations of the previously derived equation,  namely the six form equation of motion and the three form and six form duality equation respectively [4,5].  
\par
The dual graviton equation of motion that we will find involves both the dual graviton and the usual gravity field although the latter field is not present at the linearised level. As such it evades the no-go theorem of reference [11] as this assumes that the equation of motion of the dual gravity field just involves the dual gravity field. As such $E_{11}$ has provided a way to evade  the no-go theorem. 
\par
 An account of the $E_{11}\otimes_s l_1$ non-linear realisation can be found in previous papers on $E_{11}$, for example in references [4,5],  or the review of reference [6]. As is well known the fields of this theory, up to and including level three,  are the graviton $h_a{}^b$ , the three form field $A_{a_1a_2a_3}$, 
 the six  form field $A_{a_1\ldots a_6}$  and the dual graviton field $h_{a_1\ldots a_8 , b}$ respectively. The dynamics is constructed from a group element $g\in E_{11}\otimes_s l_1$ using the Cartan forms $ G _{\underline \alpha }$ and vielbein $E ^A$ which are defined  by 
$$
g^{-1} dg= E ^A l_A + G _{\underline \alpha } R^{\underline \alpha }
\eqno(1.1)$$
where $R^{\underline \alpha }$ and $l_A$ are the generators of $E_{11}$ and  the vector representation $l_1$ respectively. The Cartan forms, up to level three, are given by [4,5]
$$
G_\tau {}_{, a}{}^b = e_a{}^\rho \partial_\tau  e_\rho{}^b,
$$ 
$$
{ G}_{\tau,\mu_1 \mu_2 \mu_3} = 
e_{[a_1}{}^{\mu_1} e_{a_2}{}^{\mu_2}  e_{a_3]}{}^{\mu_3} \partial_{\tau} A_{\mu_1\mu_2\mu_3}, 
$$
$$
{ G}_{\tau , a_1\ldots a_6 }=  e_{[a_1}{}^{\mu_1}\ldots e_{a_6]}{}^{\mu_6}(\partial_{\tau} A_{\mu_1\ldots \mu_6}-A_{\mu_1\mu_2\mu_3} \partial_{\tau}A_{\mu_4\mu_5 \mu_6})
$$
$$
G_{\tau}{}_{, a_1\ldots a_8,b} =   e_{[a_1}{}^{\mu_1}\ldots e_{a_8]}{}^{\mu_8}e_{b}{}^{\nu}   (\partial_\tau h_{\mu_1\ldots \mu_8,\nu}
-A_{\mu_1\mu_2 \mu_3}\partial_\tau A_{\mu_4\mu_5\mu_6} A_{\mu_7\mu_8 \nu}
$$
$$
+2 \partial_\tau A_{\mu_1\ldots \mu_6} A_{\mu_7\mu_8 \nu}
+2\partial_\tau A_{\mu_1\ldots \mu_5 \nu} A_{\mu_6\mu_7  \mu_8})
\eqno(1.2)$$
where the vierbein is given in terms of the field $h_a{}^b$  by $e_\mu{}^a \equiv (e^h)_\mu{}^a$
\par
The advantage of working with the   Cartan  forms is that they only  transform under the Cartan involution invariant subalgebra of $E_{11}$, denoted by $I_c(E_{11})$. At level zero this subalgebra is the Lorentz algebra and the variations at the next level have the  parameter $\Lambda _{a_1 a_2  a_3 }$ under which the Cartan forms transform as   [4,5] 
$$
\delta G_{a}{}^{b}=18\, \Lambda^{c_1c_2 b }G_{c_1c_2 a}
-2\, \delta_a ^{b}  \Lambda^{c_1c_2 c_3}G_{c_1c_2 c_3},\ 
\eqno(1.3)$$
$$
\delta G_{a_1a_2a_3}=-{5!\over 2} G_{b_1b_2b_3 a_1a_2a_3}
\Lambda^{b_1b_2 b_3} -6\, G_{(c [a_1 |) } \Lambda_{c}{}_{|a_2a_3]}
\eqno(1.4)$$
$$
\delta G_{a_1\ldots a_6}=2 \Lambda_{[ a_1a_2a_3}G_{a_4a_5a_6 ]}
-112\, G_{b_1b_2b_3 [ a_1\ldots a_5,a_6]}\Lambda^{b_1b_2b_3}
+112\, G_{b_1b_2 a_1\ldots a_5a_6, b_3 }\Lambda^{b_1b_2b_3}
$$
$$
= 2 \Lambda_{[ a_1a_2a_3}G_{a_4a_5a_6 ]}
-336 \, G_{b_1b_2b_3 [ a_1\ldots a_5,a_6]}\Lambda^{b_1b_2b_3}\eqno(1.5)$$
$$
\delta G_{a_1\ldots a_8,b}=-3\, G_{[ a_1\ldots a_6}\Lambda_{a_7a_8] b}
+3 \,G_{[ a_1\ldots a_6}\Lambda_{a_7a_8 b]}
\eqno(1.6)$$
\par
The above formulae are true when the Cartan forms are written as forms, for example $G_a{}^b = dz^\Pi G_{\Pi ,}{}_a{}^b$  . However, we will work with Cartan forms for which their  first world volume index is   converted into a tangent index, that is, 
$G_A{}_{, \underline \alpha }=E_A{}^\Pi G_\Pi{}_{, \underline \alpha }$. Under the $I_c(E_{11})$ transformations this index transforms 
$$
\delta G_{a, \bullet}= -3G^{b_1b_2}{}_{,\bullet}\ 
\Lambda_{b_1b_2 a},
\quad \delta G^{a_1a_2}{}_{, \bullet}= 6\Lambda^{a_1a_2
b}  G_{b,}{}_{\bullet}, \ \ldots 
\eqno(1.7)$$
As such the Cartan forms $G_A{}_{, \underline \alpha }$ transform under equations (1.3) to (1.6) on their second ($E_{11}$) index and under equation (1.7) on their first $(l_1)$ index. 

\medskip
{\bf 2 The non-linear dual graviton equation}
\medskip
In this section we will  carry out the $E_{11}$ variation  of the six form field equation which  is second order in derivatives.In this way we will 
derive the non-linear equation for the dual graviton up to a very specific  type of term. 
We begin with the duality relation between the three form and six form  fields  which is first order in derivatives and was derived from the $E_{11}\otimes_s l_1$ non-linear realisation [3,4,5]; it takes the familiar form  
$$
E_{\mu_1\ldots \mu_7}\equiv { G}_{[\mu_1 ,\ldots \mu_7 ]}+{2\over 7 !}(\det e)^{-1}\epsilon _{\mu_1 \ldots \mu_7}{}^{\nu_1\ldots \nu_4} G_{\nu_1,\nu_2 \nu_3 \nu_4}=0
\eqno(2.1)$$
In writing this  equations we have suppressed the terms which contain derivatives with respect to the higher level coordinates; these terms can be found in references [3,4] and are also given  in the next section. 
\par
Taking the derivative of equation (2.1) we find the equation 
$$
E^{\mu_1\ldots \mu_6}\equiv \partial_{\nu} \{(\det e)^{{1\over 2}} G^{[\nu , \mu_1\ldots \mu_6]}\}=0 
\eqno(2.2)$$
which is given in terms of objects with  tangent indices by [3,4]
$$
E_{a_1\ldots a_6}\equiv (\det e)^{{1\over 2}}e^{ d \mu} \partial_{\mu} G_{[ d,a_1\ldots a_6] }
+{1\over 2}G^d{}_{,b}{}^b G_{[d,a_1\ldots a_6]}
$$
$$
-\, G^d{}_,{}^c{}_d G_{[c,a_1\ldots a_6]}-6\, G^d{}_,{}^c{}_{[a_1 |} G_{[d,c | a_2 \ldots a_6]]}=0\eqno(2.3)$$
\par
Using  the Cartan involution invariant ($I_c(E_{11})$) transformations of equations (1.3) to (1.6) we find that the variation of equation (2.3) leads to the expression. 
$$
\delta E^{a_1\ldots a_6}= -\,108\,G_{d_1,\,d_2c_1c_2}\,\Lambda^{c_1c_2[a_1|}\,G^{\left[d_1,\,d_2|a_2\,...\,a_6]\right]} + 12\,G_{d,\,c_1c_2c_3}\,\Lambda^{c_1c_2c_3}\,G^{\left[d,\,a_1\,...\,a_6\right]} 
$$
$$- 18\,G_{c_1,\,c_2c_3d}\,\Lambda^{c_1c_2c_3}\,G^{\left[d,\,a_1\,...\,a_6\right]}
$$
$$
+e_{\mu_1}{}^{[ a_1} \ldots e_{\mu_6}{}^{a_6 ]}\partial_{\nu} \{(\det e)^{{1\over 2}}
2 \Lambda^{[ \mu_1\mu_2\mu_3} G^{\nu, \mu_4\mu_5\mu_6 ]} +168\, 
(\det e)^{{1\over 2}}G^{[\nu , \mu_1\ldots \mu_6 ]\sigma_1\sigma_2 , \sigma_3}\Lambda _{\sigma_1\sigma_2  \sigma_3} \}.
\eqno(2.4)$$
In deriving equation (2.4) we have used equation of motion (2.1) and the  previously derived [4,5] three form equation of motion; these terms will be reinstated when the variation is presented at the end of this section in equation (2.11).  
\par
Using the identity 
$$
2\,\Lambda^{c_1c_2c_3}\,\varepsilon^{a_1...a_6b_1...b_5}\,S_{b_1\,...\,b_4}\,S_{ b_5 c_1c_2c_3 } - 9\,\Lambda^{c_1c_2[a_1}\,\varepsilon^{a_2\,...\,a_6]b_1...b_5 b_6}\,S_{b_1\,...\,b_4}\,S_{b_5 b_6c_1c_2}
$$
$$
= -\,{5\over 2}\,\Lambda^{[a_1a_2a_3}\,\varepsilon^{a_4a_5,a_6]b_1...b_8}\,S_{b_1\,...\,b_4}\,S_{b_5...\,b_8},
\eqno(2.5)$$
valid for any $\Lambda_{a_1a_2a_3} $ and $S_{b_1\,...\,b_4}$,  both of which are totally antisymmetric in all their indices, we find that the terms in equation (2.4) which are  quadratic in the Cartan form $G_{a_1, a_2a_3a_4}$ vanish! The remaining terms result in the equation 
$$
\delta E^{a_1\ldots a_6}=168\,e_{\mu_1}^{\,\,[a_1}...e_{\mu_6}^{\,\,a_6]}\,\partial_{\nu}\left(\left(\det{e}\right)^{{1\over 2}}\,G^{[\nu,\,\mu_1...\mu_6]\sigma_1\sigma_2,\,\sigma_3}\,\Lambda_{\sigma_1\sigma_2\sigma_3}\right)
$$
$$
=e_{\mu_1}{}^{a_1}{}\ldots e_{\mu_6}{}^{a_6}{}\{ 432\,\partial_{[\nu |} \{(\det e)^{{1\over 2}} G^{[\nu , \mu_1\ldots \mu_6\sigma_1\sigma_2]}{}_{|\tau]}       \}\Lambda _{\sigma_1 \sigma_2 }{}^{\tau}
$$
$$+216\,\partial_{\tau} \{(\det e)^{{1\over 2}} G^{[\nu , \mu_1\ldots \mu_6\sigma_1\sigma_2]}{}_{, \nu}       \}\Lambda _{\sigma_1 \sigma_2 }{}^{\tau}
$$
$$
-48\,\partial_{\nu } \{(\det e)^{{1\over 2}} G^{\sigma_1 , \sigma_2\nu\mu_1\ldots \mu_6}{}_{,\sigma_3} \Lambda _{\sigma_1 \sigma_2 }{}^{\sigma_3}\}
+864\,G^{[\nu , \mu_1\ldots \mu_6 \sigma_1  \lambda ]}{}_{,\tau } G_{\nu , (\sigma_2\lambda) } \Lambda _{\sigma_1}{}^{\sigma_2 \tau}\}
\eqno(2.6)$$
\par
As explained, for example  in reference [4],  the equation that results from the $E_{11}$ variation is computed up to terms that only contain ordinary derivatives with respect to spacetime. However, to find this result we must add certain terms to the equation that is being varied, that is in this case equation (2.3),  which contain derivatives with respect to the level one coordinate, $x^{a_1a_2}$.  The reason for this is that  any term of the form 
$f\partial_\tau g\Lambda^{\sigma_1\sigma_2 \tau}$, 
where $f$ and $g$ are any function of the fields, whose indices we have suppressed,  which  appears in the variation can be removed by the addition of a term of the form $-6f\partial^{\sigma_1\sigma_2} g$ to the equation being varied. In order not to complicate the discussion we will list these additional terms  at the end of  this section. 
\par
We observe that the second and third terms in equation (2.6) are of the form just described in the paragraph above and so we can remove these by adding appropriate terms to the $E^{a_1\ldots a_6}$ equation of motion. The last term in equation (2.6) can be processed by using the gravity-dual gravity relation of equation (3.5), derived  in the next section,  to  eliminate the dual gravity Cartan form in terms of the spin connection to find that 
$$
4\, G^{[\nu , \mu_1\ldots \mu_6 \sigma_1\lambda ]}{}_{,\tau } G_{\nu , (\sigma_2\lambda) } \Lambda _{\sigma_1}{}^{\sigma_2 \tau}
= -2\, (\det e)^{{1\over 2}}G^{[\nu , \mu_1\ldots \mu_6 \sigma_1  \lambda ]}{}_{,\tau } (\omega _{\sigma_2 ,  \nu \lambda} - Q_{\sigma_2 , \nu \lambda }) 
 \Lambda _{\sigma_1}{}^{\sigma_2 \tau}
$$
$$
=-{4\over 9 !} \epsilon ^{  \mu_1\ldots \mu_6 \sigma_1\kappa_1\kappa_2 \rho_1\rho_2} 
\omega_{\sigma_3 }{}_{,\rho_1\rho_2} \omega_{\sigma_2}{}_{, \kappa_1 \kappa_2}\Lambda_{\sigma_1}{}^{\sigma_2\sigma_3 } 
+2\, (\det e)^{{1\over 2}}G^{[\nu , \mu_1\ldots \mu_6 \sigma_1\lambda ]}{}_{\tau } Q_{\sigma_2 , \nu \lambda } 
 \Lambda _{\sigma_1}{}^{\sigma_2 \tau}
\eqno(2.7)$$
where $\omega_c{}^{ab} $ is the usual spin connection which  is given by 
$$
(\det e)^{{1\over 2}} \omega _{c, ab}= - G_{a, (bc)}+ G_{b, (ac)}+G_{c, [ab]}, \ \ {\rm and } \ \ (\det e)^{{1\over 2}} Q_a{}_{, bc}= G_a{}_{, [bc]}
\eqno(2.8)$$
The first term in  equation (2.7) vanishes due to symmetry arguments and the second terms can be removed adding terms with derivatives with respect to the level one coordinate to the equation being varied, that is, equation (2.3). 
\par
To derive  the dual gravity equation of motion we must remove the $I_c(E_{11})$ parameter $\Lambda^ {\sigma_1 \sigma_2\sigma_3}$, however, as we have just discussed,   we  can add terms to the equation that are of the form $f\partial_\tau g\Lambda^{\sigma_1\sigma_2 \tau}$  as long as we  add the corresponding term  that contain derivatives with respect to the   level one coordinates to the six form equation. Hence the dual graviton equation we derive only holds subject to the addition of such terms and is given by 
$$
E^{\mu_1\ldots \mu_8 }{}_{,\tau}\equiv \partial_{[\nu |} \{(\det e)^{{1\over 2}} G^{[\nu , \mu_1\ldots \mu_8]}{}_{,|\tau]}       \}
+\dots 
=0
\eqno(2.9)$$
where $+\ldots$ signify the possible addition of terms of the form $f\partial _\tau g$ where we have suppressed the indices $f$ and $g$ carry. The fact that it involves a derivative with respect to the $\tau$ index follows from equation (2.6)  if one observes the position of the indices on the three form parameter $\Lambda^{c_1c_2c_3}$. 
\par
The six form equation of motion  with the additional terms mentioned above is given by 
$$
{\cal E}^{a_1\ldots a_6}= e_{\mu_1}{}^{ [ a_1} \ldots e_{\mu_6}{}^{a_6 ]} \{\partial_{\nu} \{(\det e)^{{1\over 2}} G^{[\nu , \mu_1\ldots \mu_6]}\} 
-8\, \partial_\nu ( (\det e )^{{1\over2}} G^{\tau_1 \tau_2 , \nu \mu_1\ldots \mu_6 }{}_{\tau_1, \tau_2})
$$
$$
+{1\over 7} (\det e )^{-{1\over2}}
 \partial^{\mu_1\mu_2}
( (\det e )^{{1\over2}} G^{\mu_3, \mu_4 \mu_5\mu_6}) 
$$
$$-36 \, e_{\tau_1}{}^{b_1} e_{\tau_2} {}^{b_2} e_{\rho_1 b_1}e_{\rho_2 b_2 } \partial^{\rho_1\rho_2} ( (\det e )^{{1\over2}} G^{[ \nu , \mu_1\ldots \mu_6 \tau_1\tau_2 ]}{}_{ , \nu})
$$
$$
-{72} \, (\det e )^{{1\over2}} G^{[\nu , \mu_1\ldots \mu_6 \sigma_1 \lambda]} {}_{, \tau } Q^{\tau}{}_{ \sigma_1 }{}_{, \nu \lambda}\}
-3 G^{c_1c_2}{}_{,c_1c_2}{}_{ e} G^{[e , a_1\ldots a_6 ]}
$$
$$
-18\, G^{c [a_1 |} {}_{, c d_1d_2} G^{[ d_1, d_2 | a_2 \dots a_6]]} +\ldots .
\eqno(2.10)$$
where $+\ldots$ signify the addition of terms of the form $f\partial^{c_1c_2} g$ where the indices on $f$ and $g$ are suppressed. The   final variation of the six form equation of motion can be written as 
$$
\delta {\cal E}^{a_1\ldots a_6 }= 432\,\Lambda_{c_1c_2c_3}\,E^{a_1...a_6c_1c_2,\,c_3}+{8\over 7}  \Lambda ^{[a_1a_2a_3} E^{a_4a_5a_6]}
$$
$$
+{2\over 105} G_{[e_5 ,  c_1c_2c_3]} \epsilon^{a_1\ldots a_6e_1\ldots e_5} E_{e_1\ldots e_4}\Lambda^{c_1c_2c_3}
$$
$$
-{3\over 35} G_{[ e_5, e_6 c_1c_2]}\Lambda^{c_1c_2[a_1}
\epsilon^{a_2\ldots a_6 ] e_1\ldots e_6} E_{e_1\ldots e_4}
+ {1\over 420}\,\epsilon_{c_1}{}^{a_1...a_6b_1...b_4}\,\omega_{c_2,\,b_1b_2}\,E_{c_3,\,b_3b_4}\,\Lambda^{c_1c_2c_3}
\eqno(2.11)$$
where $E_{a,bc}$ is given in the next section in equation (3.50) and $E^{\mu_1\mu_2\mu_3}$ is the equation of motion of the three form gauge field which can be found by projecting equation (2.1)  and it is given by [3,4,5] 
$$
E^{\mu_1\mu_2\mu_3} \equiv \partial_{\nu} \{(\det e)^{{1\over 2}} G^{[\nu , \mu_1\mu_2 \mu_3]}\}
+
{1\over 2.4!} (\det e)^{{-1}}\epsilon ^{\mu_1\mu_2\mu_3\tau_1\ldots\tau_8} G_{[\tau_1,\tau_2\tau_3\tau_4]} G_{[\tau_5,\tau_6\tau_7\tau_8] }=0\
\eqno(2.12)$$
We use the notation ${\cal E}$ rather than $E$ when writing the symbols for the equations of motion  that contain the terms with derivatives with respect to the level one coordinates.  
\medskip
{\bf 3 The gravity-dual gravity relation }
\medskip
To find the gravity-dual gravity relation we will vary the three-six form duality relations given in equation (2.1)   under the $I_c(E_{11})$ transformations given in equations (1.3) to (1.6). However, we now use the  form of this relation when expressed in terms of four rather than seven antisymmetric indices and we include the terms which possess  derivatives with respect to the level one coordinates;
$$
{\cal  E}{}_{a_1\ldots a_4}\equiv   {\cal G}_{a_1 a_2a_3a_4 }
-{1\over 2.4!}\epsilon _{a_1a_2a_3a_4}{}^{b_1\ldots b_7}{\cal  G}_{b_1 b_2\ldots b_6, b_7 }+{1\over 2} G_{[a_1a_2}{}_{,}{}_{a_3a_4]}
\eqno(3.1)$$
where 
$$
{\cal G}_{a_1 a_2a_3a_4 }\equiv  G_{[a_1,a_2a_3a_4] }+{15\over 2}G^{b_1b_2}{}_{, b_1b_2 a_1\ldots a_4}
\eqno(3.2)$$
$$
{\cal G}_{a_1a_2 \ldots a_7 }\equiv G_{a_1,a_2 \ldots a_7 }
+28 \,G^{e_1e_2}{}_{, e_1e_2 [ a_1\ldots , a_7 ]}
\eqno(3.3)$$ 
\par
Varying under the $I_c(E_{11})$ transformations,  and using the relation  ${\cal  E}^{(1)}{}_{a_1\ldots a_4}=0$,  we find  that 
$$
e_{[ a_1}^{\mu_1}\ldots e_{a_4 ] }^{\mu_4}\delta {\cal   E}^{a_1\ldots a_4}= 
3(\det e)^{{1\over 2}} \omega_{\tau ,}{}^{\mu_1\mu_2} \Lambda^{\mu_3\mu_4\tau}
$$
$$
-{7\over 2} (\det e) ^{-{1}}\epsilon^{\mu_1\ldots \mu_4 \nu_1\ldots \nu_7}(G_{\nu_1,\nu_2\ldots \nu_7\tau_1\tau_2, \tau_3}+G_{ \tau_1, \tau_2 \tau_3 \nu_1\nu_2\ldots\nu_6 ,  \nu_7}
)\Lambda^{\tau_1\tau_2 \tau_3}=0
\eqno(3.4)$$
\par
Extracting off $\Lambda^{\rho_1\rho_2 \rho_3}$ and setting  on $\mu_3=\rho_2$ and $\mu_4= \rho_3$ and summing in equation (3.4) we find, after scaling by $(\det e )^{-{1\over 2}}$, and converting to worldvolume indices,  that 
$$
E_{\tau ,}{}^{\mu_1\mu_2}\equiv  (\det e) \omega_{\tau ,}{}^{\mu_1\mu_2} -{1\over 4} (\det e) ^{-{1\over 2}}
\epsilon ^{\mu_1\mu_2\nu_1\ldots \nu_9}G_{\nu_1 ,\nu_2\ldots \nu_9, \tau}\dot =0
$$
\centerline {or equivalently }
$$
E^{\mu_1 \mu_2 \ldots \mu_9}{}_{,\tau}\equiv (\det e) ^{{1\over 2}}G^{[\mu_1 ,\mu_2 \ldots \mu_9]}{}_{,\tau}+{2\over 9!} \epsilon ^{\mu_1 \mu_2 \ldots \mu_9\nu_1\nu_2} \omega _\tau {}_{, \nu_1\nu_2}
\eqno(3.5)$$
To find this equation we have used the identity 
$$3
X_{[a_1a_2}{}^{[c_1}\delta_{a_3a_4]}^{c_2c_3]}\delta_{c_2}^{a_3} \delta_{c_3}^{a_4} ={14\over 3} X_{a_1a_2}{}^{c_1}-{8\over 3} X_{d[a_1}{}^{d} \delta _{a_2]} ^{c_1}
\eqno(3.6)$$
for any tensor $X_{a_1a_2}{}^{c}$ which obeys $X_{a_1a_2}{}^{c}= X_{[a_1a_2]}{}^{c}$.  
\par
As we have explained previously [12,5,7,14] equation (3.5) only holds modulo certain transformations. Clearly one of these is the familiar  local Lorentz transformations which are the level zero parts of $I_c(E_{11})$. However, there are also parts of the level three gauge transformations of the field $h_{a_1\ldots a_8,b}$ and as we will discuss later in this section the gauge transformations of the three and six form gauge fields. The use of the symbol $\dot =$ signifies that the equation is to be understood in this way. 
\par
By tracing equation (3.5) we find the relation 
$$
\omega_{\rho , }{}^{\rho \mu} \dot =0
\eqno(3.7)$$
Indeed we have already used this equation in the derivation of equation (3.5). 
Clearly if this equation were to hold exactly , rather than only holding modulo the above mentioned transformations, it would be incompatible with the correct propagation of a spin two field. However, as has been explained in detail in references [12, 5, 14] once this point is taken into account  equation (3.7) is completely compatible with Einstein gravity. A gravity dual gravity relation similar to that of equation (3.5) was given in reference [3].  However, the modulo nature to the equation was only alluded to and the projection to find the gravity equation was not understood at that time and as a result it was not claimed that this was the final version of the gravity-dual gravity equation. 
\par
Given the more complicated index structure of equation (3.4) compared to equation  (3.5) it is far from clear that taking the double trace of equation (3.4) leads to the full content of  equation (3.5) and that there are not further equations which might not be compatible with the propagation of the degrees of freedom of gravity. In fact this is not the case,  
the second term on the right hand side of  equation (3.4) can be processed  as following 
$$
-\,{7\over 2}\,\left(\det{e}\right)^{-\,{1\over 2}}\epsilon^{\mu_1...\mu_4\nu_1...\nu_7}\left(G_{\nu_1,\,\nu_2...\nu_7\tau_1\tau_2,\,\tau_3} + G_{\tau_1,\,\tau_2\tau_3\nu_1...\nu_6,\,\nu_7}\right)\Lambda^{\tau_1\tau_2\tau_3} 
$$
$$
=-\,{7\over 2}\,\left(\det{e}\right)^{-\,{1\over 2}}\epsilon^{\mu_1...\mu_4\nu_1...\nu_7}\left(G_{\nu_1,\,\nu_2...\nu_7\tau_1\tau_2,\,\tau_3} + {2\over 7}\,G_{\tau_1,\,\tau_2\nu_1...\nu_7,\,\tau_3}\right)\Lambda^{\tau_1\tau_2\tau_3}
\eqno(3.8)$$
In deriving this equation  we have used the SL(11) irreducibility of the dual gravity  Cartan form, that is  $ G_{\tau,\,[ \nu_1\ldots \nu_8 , \lambda ]}=0$. We can now combine the two terms in equation (3.8) together to find the expression 
$$
-\,{1\over 2}\,\left(\det{e}\right)^{-\,{1\over 2}}\epsilon^{\mu_1...\mu_4\nu_1...\nu_7}\left(7\,G_{\nu_1,\,\nu_2...\nu_7\tau_1\tau_2,\,\tau_3} + 2\,G_{\tau_1,\,\tau_2\nu_1...\nu_7,\,\tau_3}\right)\Lambda^{\tau_1\tau_2\tau_3} 
$$
$$
=-\,{9\over 2}\,\left(\det{e}\right)^{-\,{1\over 2}}\epsilon^{\mu_1...\mu_4\nu_1...\nu_7} G_{\left[\nu_1,\,\nu_2...\nu_7\tau_1\tau_2\right],\,\tau_3}\,\Lambda^{\tau_1\tau_2\tau_3}
\eqno(3.9)$$
Finally, by using  the identity $-\,{1\over 2.9!} \epsilon_{\lambda_1\lambda_2\rho_1...\nu_9}\epsilon^{\lambda_1\lambda_2\sigma_1...\sigma_9} = \delta^{\sigma_1...\sigma_9}_{\rho_1...\rho_9}$ in  this last equation one finds the expression 
$$
{1\over 4.8!}\,\left(\det{e}\right)^{-\,{1\over 2}}\epsilon^{\mu_1...\mu_4\nu_1...\nu_7}\epsilon_{\lambda_1\lambda_2\tau_1\tau_2\nu_1...\nu_7}\epsilon^{\lambda_1\lambda_2\sigma_1...\sigma_9} G_{\sigma_1,\,\sigma_2...\sigma_9,\,\tau_3}\,\Lambda^{\tau_1\tau_2\tau_3} 
$$
$$
=-\,{3\over 4}\,\left(\det{e}\right)^{-\,{1\over 2}}\Lambda^{\tau[\mu_1\mu_2}\epsilon^{\mu_3\mu_4]\sigma_1...\sigma_9} G_{\sigma_1,\,\sigma_2...\sigma_9,\,\tau}.
\eqno(3.10)$$
\par
The result is that equation (3.4), once  we have reinstated the three form-six form relation,   can be written as 
$$
\delta{\cal E}^{\,a_1...a_4} = {1\over 4!}\,\varepsilon^{a_1...a_4b_1...b_7}\,E{}_{b_1...b_4}\,\Lambda_{b_5b_6b_7} + 3 (\det e) ^{-{1\over 2}}\,E_{c}^{[a_1a_2}\,\Lambda^{a_3a_4]c}.
\eqno(3.11)$$
From this equation it is clear that the $E_{11}$ variation of the three form-six form relation leads to the same four form equation  and the gravity-dual gravity relation of equation (3.5) and no other constraints. The derivation given here corrects a mistake in reference [7] in which  the equation (4.5) is incorrect as one of the terms was substituted by the wrong expression when writing up the paper.  
\medskip
{\bf 4 Derivation of Einstein equation from  the gravity-dual gravity relation }
\medskip
In this section we will project the gravity-dual gravity relation of equation (3.5), which is first order in derivatives,  to find an equation that is  second order in derivatives and  contains  only the graviton field.  We must also do this in such a way that the resulting equation holds exactly and so the projection must also eliminate the modular transformations which the gravity-dual gravity relation possess. We begin by considering the expression 
$$
\partial_\nu ( E_{\tau ,} {}^{\nu\mu})
-\det e \ \partial_\tau \{( \det e)^{-{1}}E_{\nu , } {}^{ab}\}  e_a{}^\mu e_b{}^\nu
\eqno(4.1)$$
where $E_{\tau ,}{}^{\mu\nu}$ is the gravity-dual gravity relation of equation (3.5). This expression would vanish if we forgot that $E_{\tau ,}{}^{\mu\nu}$ only hold modulo certain transformations. The parts of equation (4.1) that depend on $\omega_\tau^{ab}$ are 
given by 
$$
 \partial_\nu ( \det e\ \omega_{\tau ,}{}^{\nu\mu})
-\det e \ \partial_\tau \omega_{\nu , } {}^{ab}  e_a{}^\mu e_b{}^\nu = \det e R_\tau{}^\mu
\eqno(4.2)$$
The reader may verify that $R_\tau{}^a$ is the same as the well known expression for the Ricci tensor. Of course it transforms covariantly under  local Lorentz  and diffeomorphism transformations. 
\par
The part of the first term of equation (4.1) that contains the  dual graviton Cartan form is of the form 
$$
-{1\over 4} \epsilon^{\rho\mu\nu_1\ldots  \nu_9} \partial_{\rho} 
((\det e)^{-{1\over 2}} G_{\nu_1, \nu_2 \ldots \nu_9, \tau} )
\eqno(4.3)$$
We recognise that the expression on the right-hand side of this equation contains  the Bianchi identity for the dual graviton Cartan form which  can be evaluated by  considering the explicit for the Cartan form given in  equation (1.2), or by using the  $E_{11}$ Maurer-Cartan equation,  the result is  
$$
\partial_{[\rho }((\det e)^{-{1\over 2}}G_{\nu_1, \nu_2 \ldots \nu_9 ], \tau})
=
2\, (\det e)^{-{1}} (G_{[\nu_1, \nu_2 \ldots \nu_7}G_{\rho, \nu_8 \nu_9 ]\tau}
+ G_{[\nu_1, \nu_2 \ldots \nu_6 |\tau |}G_{\rho, \nu_7 \nu_8 \nu_9 ]})
\eqno(4.4)$$
\par
The part of the second term of equation (4.1) that involves the dual graviton Cartan form is then given by 
$$
{1\over 4}(\det e)^{-{1\over 2}} \partial_\tau e_\rho {}^b e_b{}^\lambda \epsilon^{\mu \rho\nu_1\ldots \nu_9}G_{\nu_1 , \nu_2\ldots \nu_9 , \lambda} 
\eqno(4.5)$$
\par
Using equations (4.4) and (4.5) we then find that 
$$
\partial_\nu ( E_\tau {}^{\nu\mu})
-\det e \ \partial_\tau \{( \det e)^{-{1}}E_{\nu , } {}^{ab}\}  e_a{}^\mu e_b{}^\nu
$$
$$
= (\det e) R_\tau {}^ \mu-4( 
12 G^{[\mu , \rho_1\ldots \rho_3 ]}G_{[\tau, \rho_1\ldots \rho_3 ]}
-\delta^{\mu }_\tau G^{[\rho_1, \rho_2\rho_3\rho_4]}G_{[\rho_1, \rho_2\rho_3\rho_4]})$$
$$
+{1\over 4} \epsilon^{\rho\mu \nu_1\ldots \nu_9}\{{2\over 3} G_{\rho ,  \nu_1\ldots \nu_6} G_{\tau , \nu_7\nu_8\nu_9} +
{1\over 3} G_{\tau , \nu_1\ldots \nu_6} G_{\rho , \nu_7\nu_8\nu_9 } 
\}
+{1\over 4} \partial_\tau e_\rho {}^b e_b{}^\lambda \epsilon^{\mu \rho\nu_1\ldots \nu_9}G_{\nu_1 , \nu_2\ldots \nu_9 , \lambda} 
\eqno(4.6)$$ 
In the second  term of this last equation we recognise the energy momentum tensor of the eleven dimensional supergravity theory. 
\par
In  equation (4.1)  we are taking the derivative of Cartan forms and as these transform in a non-trivial way one would normally have to take a covariantised derivative. However, if one  were taking the derivative of  objects that vanish then clearly no covariantisation would be required as one would be constructing an equation which is automatically true,  or put another way,  the required covariantisaion vanishes. This is the situation that occurs when we take the derivatives of the three form- six form relation of equation (2.1) to find the equation of motion for these fields. However, the  situation we encounter in equation (4.1)  is different, and indeed  rather subtle, as  the gravity-dual gravity relation of equation (3.5) does  vanish except for the fact that it  only holds modulo certain transformations. The projection we have taken is covariant with respect to local Lorentz  transformations and we now compute the two and five form gauge transformations  of the dual gravity Cartan form to find what covariantisations we must add. We take the three form and six form gauge fields to have the transformations 
$$
\delta e_\mu{}^a=0 ,\ \ \delta A_{\mu_1\mu_2\mu_3 }=\partial_{[\mu_1}\Lambda_{\mu_2\mu_3 ]} ,\ \ 
\delta A_{\mu_1\ldots \mu_6}=\partial_{[\mu_1}\Lambda_ {\mu_2\ldots \mu_6 ]} +\partial_{[\mu_1} \Lambda _{\mu_2\mu_3 }A_{\mu_3\ldots \mu_6 ]} 
\eqno(4.7)$$
These transformations are the well known gauge transformations for these fields of eleven dimensional supergravity and they leave invariant the seven form field strength $G_{[\mu_1 , \mu_2 \ldots \mu_7]}$.  They can also be deduced from the $E_{11}$ motivated gauge transformations  of reference [13]. Clearly the spin connection is invariant under theses gauge transformations. However, this is not the case for the Cartan form associated with the dual graviton field as it contains terms non-linear in the gauge fields. Using equation (4.4) we find that the object that   occurs in equation (4.3) transforms under these gauge transformations as 
$$
\delta (\partial_{[\mu_1 }((\det e)^{-{1\over 2}}G_{\mu_2, \mu_3 \ldots \mu_{10} ], \tau}))
=
-{2\over 3}  \partial_{[\mu_1} A_{\mu_2\ldots \mu_7 } \partial_{|\tau |} \partial_{ \mu_8} \Lambda _{ \mu_9 \mu_{10}]}
-{1\over 3}  \partial_{\tau} \partial_{[\mu_1}\Lambda _{\mu_2\ldots \mu_6} \partial_{\mu_7}A_{ \mu_8 \ldots \mu_{10}]}
\eqno(4.8)$$
As a result to ensure that the projection of equation (4.6) is covariant with respect to modulo transformations we  must add to the left hand side of equation (4.6) the term 
$$
-{1\over 4} \epsilon^{\nu_1\mu \nu_2\ldots \nu_{10}}\{{2\over 3} G_{\nu_1 ,  \nu_2\ldots \nu_7} G_{\tau , \nu_8\nu_9\nu_{10}} +
{1\over 3} G_{\tau , \nu_1\ldots \nu_6} G_{\nu_7 , \nu_8\nu_9\nu_{10} } 
\}
\eqno(4.9)$$
The effect of adding this term is to exactly  cancel the term quadratic in $G_{a_1 ,a_2\ldots a_4} $ in the second line on the right hand side of equation (4.6).  
\par
Preliminary calculations  indicate that by proceeding in a similar way for diffeomorphism transformations  one must add to the left hand side of equation (4.6) the term 
$$
-{1\over 4} \partial_\tau e_\rho {}^b e_b{}^\lambda \epsilon^{\mu \rho\nu_1\ldots \nu_9}G_{\nu_1 , \nu_2\ldots \nu_9 , \lambda} 
\eqno(4.10)$$
so cancelling the last term on the right hand side of equation (4.6).
\par
The effect of all these steps is the equation 
$$
\partial_\nu ( E_{\tau ,} {}^{\nu\mu})
-\det e \ \partial_\tau \{( \det e)^{-{1}})E_{\nu , } {}^{ab}\}  e_a{}^\mu e_b{}^\nu
-{1\over 4} \partial_\tau e_\rho {}^b e_b{}^\lambda \epsilon^{\mu \rho\nu_1\ldots \nu_9}G_{\nu_1 , \nu_2\ldots \nu_9 , \lambda}
$$
$$
-{1\over 4} \epsilon^{\nu_1\mu \nu_2\ldots \nu_{10}}\{{2\over 3} G_{\nu_1 ,  \nu_2\ldots \nu_7} G_{\tau , \nu_8\nu_9\nu_{10}} +
{1\over 3} G_{\tau , \nu_1\ldots \nu_6} G_{\nu_7 , \nu_8\nu_9\nu_{10} } 
\}
$$
$$
=0= (\det e)R_{\tau }{}^\mu - 
4( 12  \, G^{[\mu , \rho_1\ldots \rho_3 ]}G_{[\tau, \rho_1\ldots \rho_3 ]}
-\delta^{\mu }_\tau G^{[\rho_1, \rho_2\rho_3\rho_4]}G_{[\rho_1, \rho_2\rho_3\rho_4]})
\eqno(4.11)$$
We recognise the last equation as the equation of motion for gravity of eleven dimensional supergravity. This equation has previously been derived as following from the $E_{11}\otimes_s l_1$ non-linear realisation but by varying the three form equation of motion, which is second order in derivatives,  under $I_c(E_{11})$ transformations [4,5]. In this later way of proceeding one does not encounter any equations that are modulo and as a result it confirms that the covariantisation of modulo transformations used above is consistent with $E_{11}$. 
\medskip
{\bf {Discussion}}
\medskip
In this paper we have computed the  gravity and dual gravity relation which is first order in derivatives and follows from the $E_{11}\otimes_sl_1$ non-linear realisation, that is E theory. By varying the previously deduced six form equation of motion we have found  dual gravity equation of motion that is second order in derivatives up to the presence of some  very specific  terms whose form is  discussed in the paper.  These results, together with the linearised results up to level four,  should leave the reader in no doubt that the conjecture that E theory contains eleven dimensional supergravity,   given in references [1,2] has been demonstrated. The equations of motion follow  essentially uniquely  from the Dynkin diagram of $E_{11}$. 
\par
E theory is a unified theory in that it contains not only eleven dimensional supergravity but also all the maximal supergravities, depending what decomposition one takes,  and also all the gauged maximal supergravities, depending which  fields one turns on, see reference [6] for a review. As such E theory  should  be the low energy effective action for type II strings and branes replacing the maximal supergravities in this role. There are many more coordinates and fields in E theory beyond those found in supergravity and it would be good to systematically understand what their physical meaning is. 
\par
In the revised version of this paper we have corrected the derivation of the dual graviton equation to make it clear that it was only derived up to the some very specific terms. The calculation to fix these terms given in the earlier version of the paper was incorrect as it wrongly  assumed that the dual graviton transformed under diffeomorphisms as if it was a standard general relativity tensor.  It was only in the earlier version of this paper that  an argument outside the framework of $E_{11}$ was resorted to. Recently the full dual graviton equation has been derived from  the variation of the six form equation of motion given  in section two by requiring  that the dual graviton equation have the same symmetries as the dual graviton field [15]. This derivation is entirely in the context of the $E_{11}\otimes_s l_1$ non-linear realisation. 

\medskip
{\bf {Acknowledgements}}
\medskip
We wish to thank Nikolay Gromov for help with the derivation of the equations of motion from the non-linear realisation. We also wish to thank the SFTC for support from Consolidated grant number ST/J002798/1 and Alexander Tumanov wishes to thank King's College  for the support they provided by his  Graduate School International Research Studentship,  the 
ÊIsrael Science Foundation (grant number 968/15)Ê and  CERN Theoretical Physics Division.  
\medskip
{\bf {References}}
\medskip
\item{[1]} P. West, {\it $E_{11}$ and M Theory}, Class. Quant.  
Grav.  {\bf 18}, (2001) 4443, hep-th/ 0104081. 
\item{[2]} P. West, {\it $E_{11}$, SL(32) and Central Charges},
Phys. Lett. {\bf B 575} (2003) 333-342,  hep-th/0307098. 
\item {[3]} P. West, {\it Generalised Geometry, eleven dimensions
and $E_{11}$}, JHEP 1202 (2012) 018, arXiv:1111.1642. 
\item {[4]} A. Tumanov and P. West, {\it E11 must be a symmetry of strings and branes},  arXiv:1512.01644. 
\item{[5]} A. Tumanov and P. West, {\it E11 in 11D}, Phys.Lett. B758 (2016) 278, arXiv:1601.03974. 
\item{[6]}  P. West, {\it Introduction to Strings and Branes}, Cambridge University Press, 2012. 
 \item{[7]}  A. Tumanov and and P. West, {\it $E_{11}$,  Romans theory and higher level duality relations}, IJMPA, {\bf Vol 32}, No 26 (2017) 1750023,  arXiv:1611.03369.  
\item{[8]} T. Curtright, {\it Generalised Gauge fields}, Phys. Lett. {\bf 165B} (1985) 304. 
\item{[9]} C. Hull, {\it   Strongly Coupled Gravity and Duality}, Nucl.Phys. {\bf B583} (2000) 237, hep-th/0004195. 
\item{[10]} N.  Boulanger, S. Cnockaert  and  M.  Henneaux, {\it A note on spin-s duality}, JHEP 0306 (2003) 060, hep-th/0306023. 
\item{[11]} X. Bekaert, N. Boulanger and M. Henneaux, {\it Consistent deformations of dual formulations of linearized gravity: A no-go result } 
Phys.Rev. D67 (2003) 044010,  arXiv:hep-th/0210278. 
X.  Bekaert, N.  Boulanger and  S.  Cnockaert, {\it No Self-Interaction for Two-Column Massless Fields}, J.Math.Phys. 46 (2005) 012303, arXiv:hep-th/0407102. 
\item{[12]} P. West, {\it Dual gravity and E11},  arXiv:1411.0920.
\item{[13]} P. West, {\it  Generalised Space-time and Gauge Transformations}, JHEP 1408 (2014) 050, arXiv:1403.6395. 
\item{[14]} P. West, {\it On the different formulations of the E11 equations of motion }, Mod.Phys.Lett. A32 (2017) no.18, 1750096, arXiv:1704.00580. 
\item{[15]} K. Glennon and P. West, {\it The non-linear dual gravity equation of motion in eleven dimensions }, arXiv:2006.02383.

\end